\documentclass[]{spie}  

\usepackage{amsmath,amsfonts,amssymb}
\usepackage{graphicx}
\usepackage{color}
\usepackage[colorlinks=true, allcolors=blue]{hyperref}

\title{Simons Observatory: Observatory Scheduler and Automated Data Processing}

\author[a,b]{Yilun Guan}
\author[c,d]{Kathleen Harrington}
\author[e]{Jack Lashner}
\author[e]{Sanah Bhimani}
\author[f]{Kevin T. Crowley}
\author[j,k]{Nicholas Galitzki}
\author[g]{Ken Ganga}
\author[i]{Matthew Hasselfield}
\author[b,h]{Adam D. Hincks}
\author[f]{Brian Keating}
\author[e]{Brian J. Koopman}
\author[l]{Laura Newburgh}
\author[e]{David V. Nguyen}
\author[e]{Max Silva-Feaver}
\affil[a]{Dunlap Institute for Astronomy \& Astrophysics, University of Toronto, 50 St. George St., Toronto ON M5S 3H4, Canada}
\affil[b]{David A. Dunlap Department of Astronomy and Astrophysics, University of Toronto, 50 St. George St., Toronto ON M5S 3H4, Canada}
\affil[c]{High Energy Physics Division, Argonne National Laboratory, Lemont, IL 60439, USA}
\affil[d]{Department of Astronomy and Astrophysics, University of Chicago, Chicago, IL, 60637, USA}
\affil[e]{Wright Laboratory, Department of Physics, Yale University, New Haven, CT 06520, USA}
\affil[f]{Department of Physics, UC San Diego, La Jolla, CA 92093 USA}
\affil[g]{Université Paris Cité, CNRS, Astroparticule et Cosmologie, F-75013 Paris, France}
\affil[h]{Specola Vaticana (Vatican Observatory), V-00120 Vatican City State}
\affil[i]{Center for Computational Astrophysics, Flatiron Institute, 162 Fifth Avenue, New York NY, 10010}
\affil[j]{Department of Physics, University of Texas at Austin, Austin, TX, 78712, USA}
\affil[k]{Weinberg Institute for Theoretical Physics, Texas Center for Cosmology and Astroparticle Physics, Austin, TX 78712, USA}
\affil[l]{Department of Physics, Yale University, New Haven, Connecticut 06520, USA}

\authorinfo{Email: yilun.guan@utoronto.ca}

\makeatletter
\let\jnl@style=\rmfamily 
\def\ref@jnl#1{{\jnl@style#1}}%
\newcommand\apj{\ref@jnl{ApJ}}
\newcommand\apjs{\ref@jnl{ApJS}}
\newcommand\jcap{\ref@jnl{JCAP}}

\makeatother

\begin{document} 

\maketitle

\begin{abstract}
The Simons Observatory (SO) is a next-generation ground-based telescope located in the Atacama Desert in Chile, designed to map the cosmic microwave background (CMB) with unprecedented precision. The observatory consists of three small aperture telescopes (SATs) and one large aperture telescope (LAT), each optimized for distinct but complementary scientific goals. To achieve these goals, optimized scan strategies have been defined for both the SATs and LAT. This paper describes a software system deployed in SO that effectively translates high-level scan strategies into realistic observing scripts executable by the telescope, taking into account realistic observational constraints. The data volume of SO also necessitates a scalable software infrastructure to support its daily data processing needs. This paper also outlines an automated workflow system for managing data packaging and daily data reduction at the site.
\end{abstract}


\section{INTRODUCTION}
\label{sec:intro}
Over the past few decades, significant experimental progress has been made in measuring the cosmic microwave background (CMB), the oldest light in the universe. This progress has led to an era of precision cosmology, with experiments such as \textit{Planck} \cite{Planck2013:XVI:CP}, the Atacama Cosmology Telescope (ACT) \cite{ACT:Aiola:2020}, the South Pole Telescope (SPT) \cite{SPT3G:2014}, and the BICEP/Keck experiments \cite{BK:Ade:2022:BICEP3} achieving unprecedented precision in measuring CMB anisotropies at the level of ten parts per million. These measurements have been the cornerstone of our standard cosmological model, the $\Lambda$CDM model, and have placed percent-level constraints on the fundamental properties of our universe. However, key questions remain unanswered, including the existence of primordial gravitational waves from cosmic inflation and the nature of dark matter and dark energy. To address these questions, next-generation CMB experiments are being built that will offer significantly improved sensitivity in CMB polarization measurements at both small and large angular scales.

The Simons Observatory (SO) is one of the next-generation experiments that will push the frontier in precision measurements of the CMB \cite{SO:2019:SciGoal}. Located in the Atacama Desert at an altitude of 5,200 meters in Chile's Parque Astronomico, the SO consists of one large aperture telescope (LAT) with a 6-meter primary mirror, and three 0.5-meter small aperture telescopes (SATs). This combination of SATs and LAT enables the SO to make precise measurements of the CMB anisotropies from degree-scale to arcminute-scale. Specifically, the LAT receiver is equipped with around 30,000 transition-edge sensor (TES) bolometric detectors, distributed across seven optics tubes. Each optics tube has three detector modules, each contains a detector wafer that hosts a hexagonal assembly of detector arrays sensitive to two frequency bands. Thanks to the large number of detectors, the LAT can map the CMB sky with arcminute resolution and an approximately 7.8$^\circ$ field of view. In comparison, the mid-frequency (MF; 90/150~GHz) and ultra-high-frequency (UHF; 220/280~GHz) SATs each has around 12,300 detectors installed on seven wafers, fitted in a single optics tube. SAT also features a rotating cryogenic half-wave plate (HWP), which modulates the polarization of incoming light. SAT maps the sky at 0.5$^\circ$ resolution at 93~GHz with a total field of view of around 35$^\circ$ \cite{Galitzki:2024}.

The success of the SO in achieving its science goals depends crucially on the development of an optimized scan strategy and observing program \cite{Stevens:2018}. For SATs, the primary objective is to detect the faint primordial gravitational wave signal in the B-mode polarization, which requires a dedicated scanning strategy that optimizes for noise statistics and foreground cleaning. To achieve this objective, SATs will employ a deep scanning strategy, focusing on small patches of sky with low galactic foreground contamination. In comparison, the LAT will prioritize having large sky coverage, especially in areas that overlap with external surveys such as the Vera C. Rubin Observatory \cite{LSST:2009:SB} and the Dark Energy Spectroscopic Instrument (DESI) \cite{DESI:2016:I}. This strategy will allow the LAT to effectively probe small-scale anisotropies, including CMB lensing and galaxy clusters, capitalize on joint analysis with these external surveys \cite{DeBernardis:2016}, and open doors to Galactic science and time-domain science.

With scan strategies defined, the next essential step is to transform the theoretical scan strategies into detailed and executable observing scripts for the telescopes. This is the primary function of the observatory scheduler, responsible for not only effectively executing the main scan strategies but also incorporating a variety of operational requirements essential for instrument health and characterization. These requirements can include scheduling calibration scans using planets, conducting routine IV curve and bias step measurements,\footnote{The IV curve, also known as current-voltage relation, is a key measurement that characterizes the superconducting transition of the TES. It measures the current response of the TES as we vary the voltage applied via the bias lines. The bias step is an alternative method to characterize detectors by measuring the detector's response to a small-amplitude square-wave added to the bias lines. It measures key calibration parameters such as the resistance of TES, $R_{\rm TES}$, responsivity, and time constants of the TES.} and managing the operations of the HWP. By seamlessly integrating the main scan strategy with the operational needs of the instruments, the observatory scheduler is key to the effective execution of an observing plan, thereby maximizing its observing efficiency. We will provide an overview of the observatory scheduler system in the first part of this paper.

In addition to the observatory scheduler, another critical component of the Simons Observatory's software infrastructure is the automated workflow system, which is responsible for automating routine data processing tasks. Traditional automated workflow systems, such as \texttt{cronjobs}, are commonly used for scheduling script executions at specified times. However, these systems often lack transparency, making it difficult to monitor execution status, inspect logs, and manage task dependencies and concurrency. Furthermore, traditional workflow systems also lack flexibility in controlling execution and its environment.

To overcome these limitations, more modern and robust workflow management tools like Apache Airflow\footnote{\url{https://airflow.apache.org/}} and \texttt{Prefect}\footnote{\url{https://www.prefect.io/}} have been invented. These tools are capable of providing a dynamic, extensible, and scalable architecture to manage workflows as directed acyclic graphs of tasks. They also include rich command-line utilities, user-friendly web interface for monitoring task progress, managing dependencies, and much more, thereby offering a comprehensive and flexible solution for automating complex data processing workflows. SO has adopted \texttt{Prefect} to coordinate its daily data processing. \texttt{Prefect} stands out because it allows the definition of workflow in Python, which fits naturally with the SO software stack. The Python support also makes it easier to transform the data processing pipelines in SO, as part of the \texttt{sotodlib} library,\footnote{\url{https://github.com/simonsobs/sotodlib}} into automated workflows managed by \texttt{Prefect}.

Within SO, this \texttt{Prefect}-based automated workflow system plays the pivotal role of automating daily data packaging and data reduction pipelines running at the observatory site. The daily data packaging pipeline, specifically, is the critical first step in data analysis at the site which re-bundles low-level data from raw acquisition format into a performance-optimized data structure and resamples relevant housekeeping data to synchronize with the detector timestream to facilitate easier and more efficient downstream data analysis. The workflow system also manages daily data reduction tasks from data preprocessing to, eventually, daily mapmaking, an important prerequisite to the daily transient discovery program in SO. In the second part of this paper we will describe this workflow system in SO and its application in automating data packaging and data reduction pipelines.

This paper will be structured as the following. Section~\ref{sec:scheduler} first discusses the general design of the observatory scheduler. Section~\ref{sec:workflow} describes the application of the \texttt{Prefect}-based automated workflow system in SO. We conclude in Section~\ref{sec:conclusion}.

\section{Observatory Scheduler}
\label{sec:scheduler}
Observation planning in SO is a multi-level process designed to
maximize observing efficiency and optimize the respective scientific
goals of different instruments. At the highest level, we define
different scan strategies for the SATs and the LAT, respectively.

The observatory scheduler is an integral part of the observation
planning system in SO, being responsible for transforming a scan
strategy into a detailed and executable observing script for the
instrument. The input to the observatory scheduler is a so-called
\textit{master schedule} that is defined based on the higher-level
scan strategy. The master schedule is composed of a set of scan
blocks, each of which contains constant elevation scans of a specific
sky patch. In SO, we can produce this master schedule using either a
``classical'' scheduler, or an ``opportunistic''
scheduler~\cite{Stevens:2018}. The classical scheduler, which will be
used during commissioning and early observations, typically
orchestrates observations simply by cycling through a pre-defined list
of target fields, during both periods when the field is
rising and setting in the sky. The opportunistic scheduler on the
other hand, like that used in Advanced ACTPol \cite{DeBernardis:2016},
sub-divides target fields into tiles and dynamically adjusts
priorities based on the frequency of visits to each tile, after
accounting for observational constraints like Sun avoidance or Moon
avoidance. This can be used later for SO, when the various possible
systematics which might affect the observations are understood.
Table~\ref{tab:example master schedule} shows an example master
schedule, where each row corresponds to an approximately one-hour
period of constant elevation scan, parameterized by the target
azimuth, elevation, boresight rotation, and the azimuth range of the
scan.

\begin{table}[t]
\caption{Example master schedule for the SAT for illustration. Rotation specifies the boresight rotation applied in units of degrees. Az Min/Max specify the boundary of the azimuth scan in units of degrees.}
\begin{tabular}{|c|c|c|c|c|c|c|c|}\hline
Start Time (UTC) & Stop Time (UTC) & Rotation [$^\circ$] & Patch & Az Min [$^\circ$] & Az Max [$^\circ$] & Elevation [$^\circ$] \\\hline
2024-01-01 20:35:00 & 2024-01-01 21:40:00 & -45.00 & Patch 1 & 105.53 & 145.53 & 50.00 \\
2024-01-01 21:40:00 & 2024-01-01 22:45:00 & -45.00 & Patch 1 & 105.53 & 145.53 & 50.00 \\
2024-01-01 22:45:00 & 2024-01-01 23:50:00 & -45.00 & Patch 1 & 105.53 & 145.53 & 50.00 \\
2024-01-01 23:50:00 & 2024-01-02 00:55:00 & -45.00 & Patch 2 & 214.47 & 254.47 & 50.00 \\
2024-01-02 00:55:00 & 2024-01-02 02:00:00 & 0.00 &  Patch 2 & 214.47 & 254.47 & 50.00 \\
2024-01-02 02:00:00 & 2024-01-02 03:05:00 & 0.00 &  Patch 2 & 214.47 & 254.47 & 50.00 \\
... & ... & ... & ... & ... & ... & ... \\
\end{tabular}
\label{tab:example master schedule}
\end{table}

\begin{figure}[t]
\centering
\includegraphics[width=\textwidth]{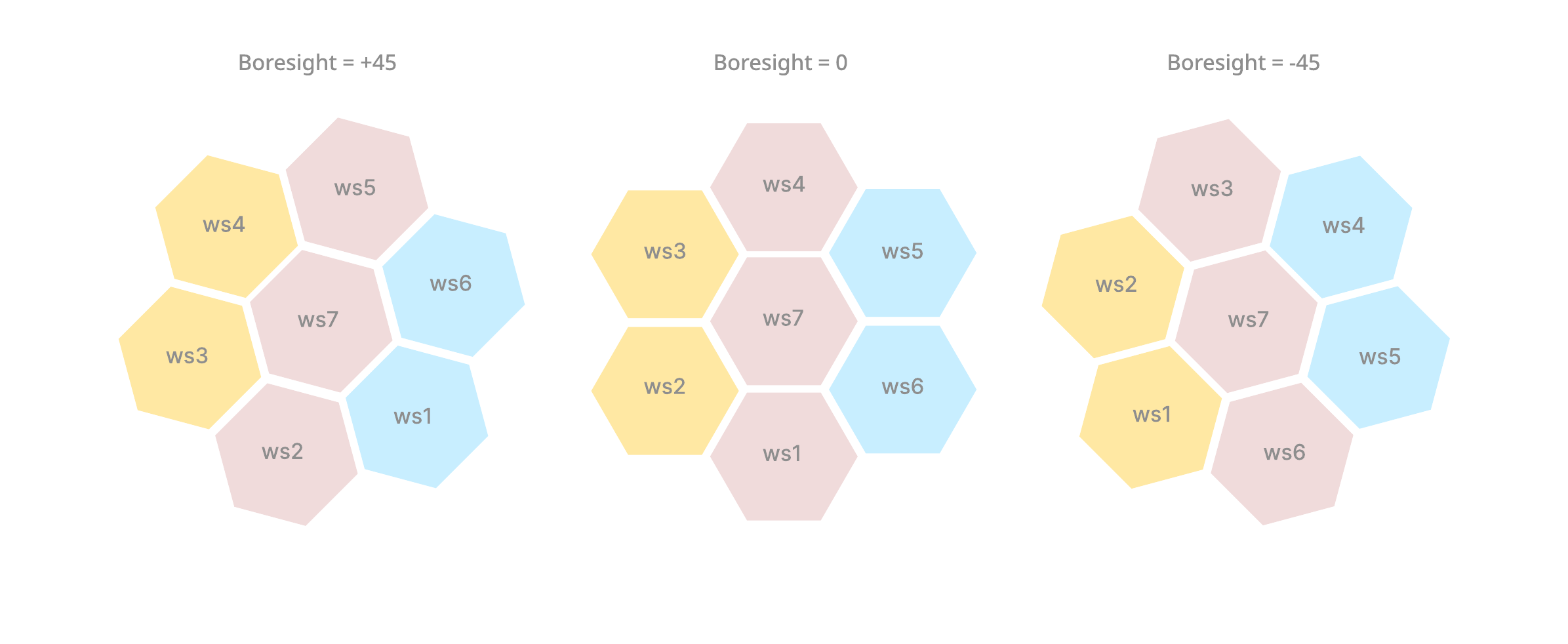}
\caption{Definition of wafer groups on the SAT focal plane with boresight rotation set to $+45^\circ$, $0^\circ$, and $-45^\circ$, respectively. Three wafer groups are defined as approximately vertical columns, color coded as yellow, red, and blue, respectively. \texttt{ws\#} denotes a specific wafer slot, where ``wafer slot'' is a unique identifier for a sub-array location in the full SAT focal plane.}
\label{fig:wafer groups}
\end{figure}

Using the master schedule as input, the scheduler first constructs a
daily observing plan that adheres to the master schedule while
incorporating additional operational requirements, such as the
inclusion of calibration targets essential for determining the
pointing offsets of detectors and their relative optical gain. For
instance, the SATs require routine calibration using Jupiter and
Saturn observations. The scheduler injects these planet targets into
the observing plan on a day-to-day basis, depending on the calibration
strategy. The entire field-of-view over 30 degrees, which is too large to efficiently scan with sufficient signal-to-noise during a calibration operation. Instead we sub-divide the full
focal plane into separate wafer groups, each containing a column of
wafers. Figure~\ref{fig:wafer groups} shows the definition of three
wafer groups in the SAT focal plane, at different boresight rotation
settings. Wafer groups are typically defined as an approximately
vertical column of wafers to minimize the azimuth throw required to
cover all detectors in the wafer group. The scheduler supports
scheduling planet scans targeting a specific wafer group, thereby
improving the scan density of each planet observation and the
calibration quality. We discuss the detailed algorithm of planet scan
planning in Section~\ref{sec:planet}.

\begin{figure}[t]
\centering
\includegraphics[width=0.6\textwidth]{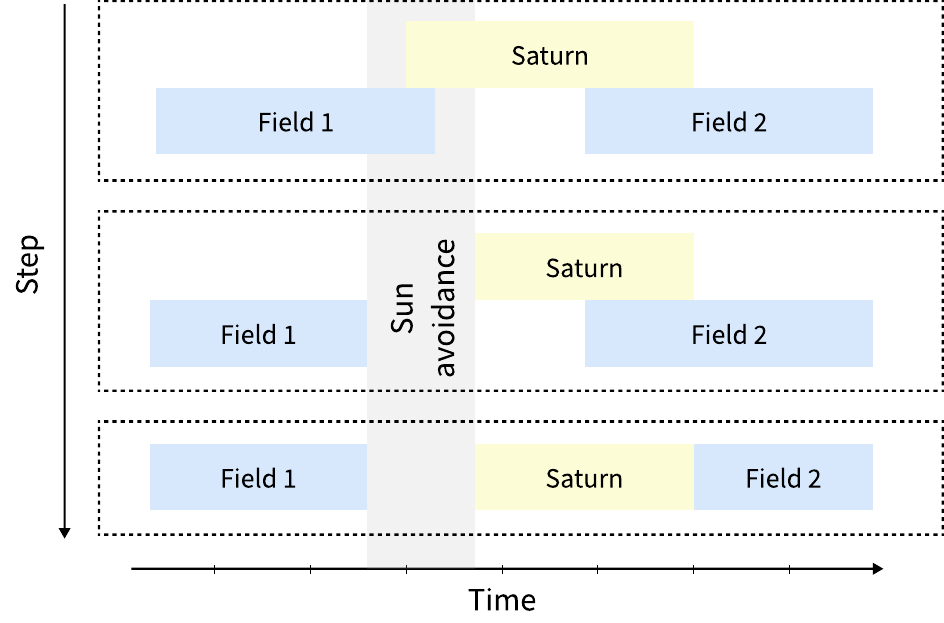}
\caption{An example observing planning process. In panel 1 we inject a Saturn scan in the master schedule; in panel 2 we trim both the CMB and the planet scans based on the sun avoidance rule; in panel 3 we resolve the scheduling conflict between Saturn scan and cmb scan by prioritizing calibration.}
\label{fig:conflict}
\end{figure}

In addition to adding planet scans, the scheduler also takes into
account various other operational constraints set by the telescope
operator. These constraints are defined as \textit{rules} that modify
each scan block accordingly. One of the most critical rules is the
\textit{sun avoidance} rule, which trims scan blocks so that they do
not reach within a pre-defined radius around the Sun, thereby
preventing instrument damage. Figure~\ref{fig:conflict} shows an
example of a sun avoidance rule being applied to trim both CMB and
planet scan blocks. We discuss sun avoidance rules in details in
Section~\ref{sec:sun avoidance}. To facilitate day-to-day operation
requirements in an observing day, the scheduler implements a suite of
rules, in addition to sun avoidance, that can be turned on when
planning observations, such as the azimuth range rule which trims each
scan to within a specified azimuth range.

When two scan blocks overlap in time, conflict resolution becomes
necessary. In the early stages of SO, we adopt a prioritization policy
that favors calibration targets over CMB scans. As a result, CMB scan
blocks may be trimmed or split to accommodate calibration scans.
However, as our understanding of the instrument evolves and as we
start to maximize science outputs, we anticipate this prioritization
policy to adapt and change. In the case when two calibration targets
conflict in time, the lower priority target, as defined by the remote
observer, will be skipped. The skipped calibration target will get a
higher priority in the next planet passage. Figure~\ref{fig:conflict}
shows an example of conflict resolution where the CMB scan is trimmed
to prioritize a Saturn scan.

With the observing plan established, the scheduler's next step is to
break down the plan into a detailed list of operations. These
operations include setup procedures at the beginning of an observing
session, termed \textit{pre-session} operations, wrap-up procedures at
the end of an observing session, termed \textit{post-session}
operations, operation sequences that set up the telescope before each
scan, such as spinning up the half-wave plate when observing a planet,
termed \textit{pre-scan} operations, and follow-up tasks after each
scan, such as bias step measurements, termed \textit{post-scan}
operations. To observe the targets on schedule, the scheduler must
fully account for the timing of pre- and post-scan operation
sequences, which is especially critical for planet observations due to
the limited observing window to observe a planet at a given elevation.
Planning operations is a non-trivial problem, as it requires careful
accounting of various state dependencies and operational constraints.
The SO scheduler adopts a multi-pass approach to plan for operations,
which we discuss in more detail in Section~\ref{sec:operation}.

When scheduling pre- and post-scan operations, it is crucial to ensure
the telescope's safety from the Sun during these operations, which
leads to an additional scheduling constraint. A traditional approach
would typically restrict these operations to a sun-safe interval, but
this may result in the unnecessary loss of valuable observation time.
To overcome this issue, the scheduler focuses on identifying
alternative sun-safe azimuth positions for the telescope during the
pre-scan operations. By selecting a sun-safe azimuth for the duration
of the operations, the telescope can maintain a constant elevation for
scanning while staying clear of the Sun. Once the pre-scan phase
concludes, the telescope moves azimuthally to the desired observing
pointing at the scheduled time. This revised strategy improves the
total observation time when pre-scan operations are affected by Sun
exposure. A more detailed discussion of this approach can be found in
Section~\ref{sec:azimuth}.

After establishing the operation plan, the scheduler will translate
each operation into a series of Python commands based on the
\texttt{sorunlib}\footnote{\url{https://github.com/simonsobs/sorunlib}}
Python library, which interfaces with the Observatory Control
System\cite{Koopman:2024}. Like the operation planning process, the
commands generated may be state-dependent. This translation step,
therefore, also accounts for the state progression to generate
commands that accurately reflect the telescope's operational states at
various times. In addition, the commands generated for each operation
are interspersed with \texttt{wait} commands, which synchronize the
execution of the script with the given operation schedule. The result
of this step is an executable Python script composed of
\texttt{sorunlib} commands. This script is then passed to and executed
by the sequencer, \texttt{nextline},\cite{Sakuma:2020:Nextline} an
online Python interpreter connected to the observatory control system
to facilitate the live control and operation of the observatory.

\begin{figure}[t]
  \centering
  \includegraphics[width=0.6\textwidth]{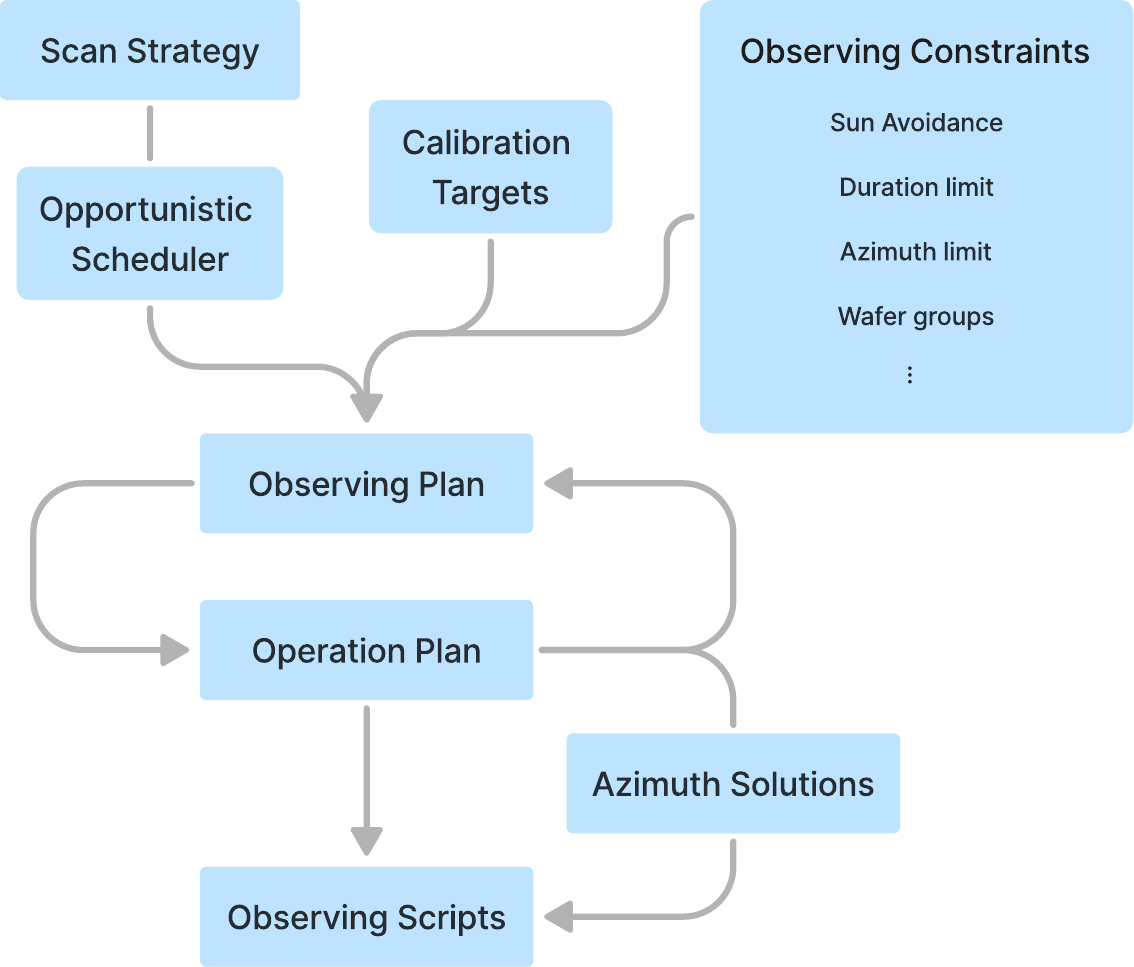}
  \caption{Overview of the scheduler's workflow.}
  \label{fig:scheduler_overview}
\end{figure}

Figure~\ref{fig:scheduler_overview} shows an overview of the
scheduling algorithm described above. We refer to this algorithm as
our baseline scheduling program. In practice, alternative scheduling
programs may be necessary to perform instrument tests or special
observations. To support flexibility in switching between different
scheduling programs, we maintain a database that associates each
program with a specific date range. The sequencer, \texttt{nextline},
supports automatic retrieval of a new observing script upon completion
of a running one by sending a request to the scheduler. The scheduler
then references the database of observing programs to identify an
active program for the date requested and redirects the request to the
relevant scheduling program to produce an observing script and return
to \texttt{nextline}. This approach has proven particularly useful in
the early stages of SO when auto-generated observing scripts are often
interspersed with manually-developed scripts by the telescope
operators to perform special instrument tests.

\subsection{Planet scan}\label{sec:planet}
Planet scans are injected into the master observing plan on a
day-to-day basis based on instrumentation requirements. With the large
focal plane sizes in SO and the short observing window of a planet
transit, we typically divide the focal plane into several groups of
wafers (wafer groups) and set each planet scan to target one specific
wafer group. Figure~\ref{fig:wafer groups} shows the definition of
three wafer groups in the SAT focal plane, as three vertical columns.
Such a vertical split minimizes the azimuth throw needed to cover all
the detectors, thus improving the scan density and calibration
quality. With each planet scan focusing on one wafer group, we need
multiple planet transits to fully map the focal plane. This typically
takes observations over several days to finish. We define calibration
targets as the combination of a planet target, the wafer group,
elevation, and, for the case of SAT, the boresight rotations.
Table~\ref{tab:caltarget examples} gives an example of a list of
calibration targets that can be inserted to the master observing plan.

\begin{table}[h]
\centering
\caption{Example calibration targets for SAT: ws\# stands for specific wafer slot. Rotation refers to boresight rotation to be applied.}
\begin{tabular}{|c|c|c|c|}
\hline
Source & Wafer Group & Elevation & Rotation \\
\hline
Jupiter & ws1, ws2 & $50^\circ$ & $0^\circ$ \\
\hline
Jupiter & ws2, ws3 & $50^\circ$ & $45^\circ$ \\
\hline
Saturn & ws2, ws3 & $50^\circ$ & $-45^\circ$ \\
\hline
\end{tabular}
\label{tab:caltarget examples}
\end{table}

\begin{figure}[t]
  \centering
  \includegraphics[width=0.7\textwidth]{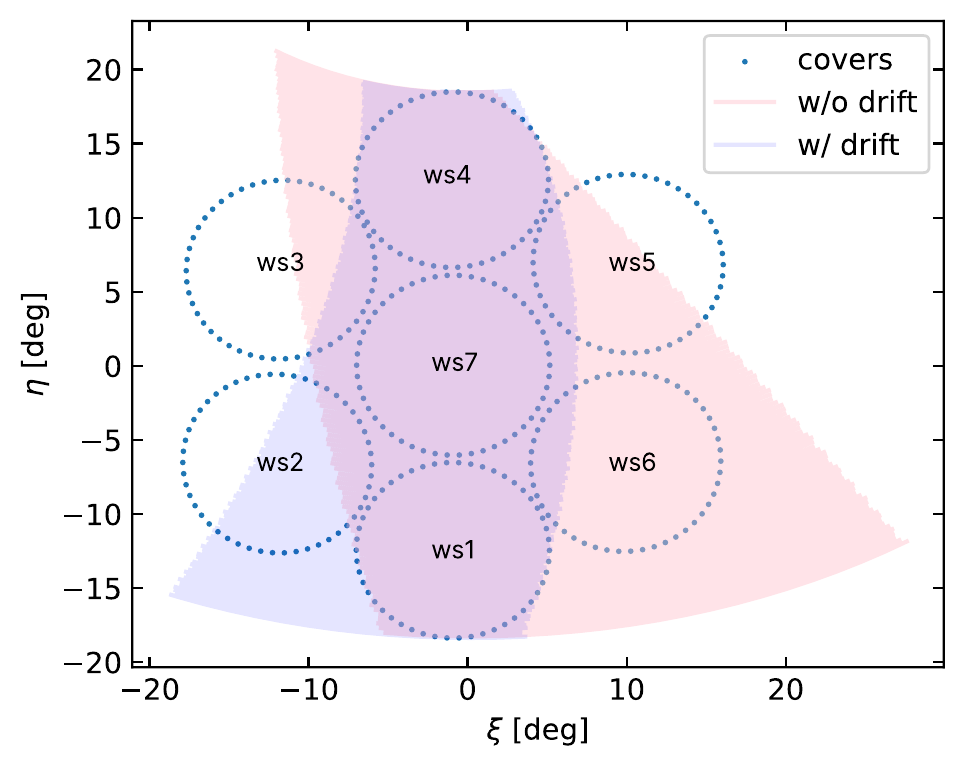}
  \caption{An example Saturn rising scan that targets the central column of wafers (wafer slots: \texttt{ws1}, \texttt{ws4}, \texttt{ws7}). The blue dots represent the detector cover for different wafers. The shaded area represents the trajectory of the planet scan solution in focal plane coordinates. The shaded area in red represents the planet solution without enabling azimuth drift; the shaded area in blue represents that with azimuth drift enabled. When azimuth drift is enabled, the azimuth throw required to cover all detectors in the target wafer group is significantly reduced, from 40.8$^{\circ}$ to 27.0$^{\circ}$.}
  \label{fig:drift}
\end{figure}

For a given calibration target, we solve for an optimal scan at a
target elevation, el$_0$, that minimizes the total observation time
needed to cover every detector in the target wafer group with the
following steps.
\begin{itemize}
  \item We first pre-compute the trajectory of the source, az$_s(t)$,
        el$_s(t)$, for the full time span of the observing plan when
        the source is above the horizon and split the full observing
        window based on rising or setting modes. Here az and el
        represent azimuth and elevation, respectively.
  \item We parametrize each wafer slot in the target wafer group as a
        circle in boresight coordinate, $(\xi, \eta)$,\footnote{The
        coordinate, $\xi=\eta=0$, corresponds to the boresight center}
        and produce dummy detectors that trace the outer boundary of
        the wafer group, with coordinates $(\xi_i, \eta_i)$ for the
        $i$-th dummy detector. These dummy detectors allow for easy
        tracking of the apparent geometry changes of each wafer in
        horizontal coordinates (az, el) at different points on the
        sky. We refer to these dummy detectors as ``detector cover''
        hereafter.
  \item We compute the center of the target wafer group as the average
        of the detector cover, $(\xi_w, \eta_w)$, and solve for an
        azimuth az$_0$ such that when our boresight is at (az$_0$,
        el$_0$), the center of the wafer group, $(\xi_w, \eta_w)$,
        falls on the trajectory of the source, (az$_s(t)$, el$_s(t)$).
        We then calculate the sky coordinates of each detector in the
        detector cover, (az$_i$, el$_i$) when our boresight is at
        (az$_0$, el$_0$). This step effectively moves the boresight to
        the target elevation such that the center of the wafer group
        approximately intersects with the source trajectory.
  \item We can then easily design an optimal scan for this calibration
        target. For a rising source, the start time of the scan is,
        $t_0 = \{t~\vert~{\rm el}_s(t_0) = \min\{{\rm el}_i\}\}$, and
        the stop time is
        $t_1 = \{t~\vert~{\rm el}_s(t_1) = \max\{{\rm el}_i\}\}$. The
        azimuth throw of the scan will go from
        ${\rm az}_{\min} = \min\{{\rm az}_i\}$ to
        ${\rm az}_{\max} = \max\{{\rm az}_i\}$. A setting scan can be
        planned similarly with the start and stop times reversed.
  \item SO also supports adding a constant azimuth drift to the scan,
        introducing $\Delta{\rm az}(t) = v_{\rm drift} t$, with
        $v_{\rm drift}$ a constant drift speed. This has the potential
        to further reduce the azimuth throw required to cover a target
        wafer group to improve the observing efficiency and scan
        density. To build an optimal scan with azimuth drift enabled,
        we first compute the effective trajectory of the source by
        subtracting the effect of drift,
        $({\rm az}_{\rm eff, s}, {\rm el}_{\rm eff, s}) = ({\rm az}_s - \Delta {\rm az}, {\rm el}_s)$,
        and then perform the same steps as above as if no drift is
        applied. We repeat this process for different $v_{\rm drift}$
        to identify the optimal $v_{\rm drift}$ that minimizes the
        azimuth throw of the scan using a \texttt{scipy} optimizer.
        Figure~\ref{fig:drift} shows a comparison between a planet
        scan solution with azimuth drift enabled and one without for a
        rising Saturn targeting the central column of wafers. In both
        solutions, the planet will scan across all detectors in the
        target wafer group, but with drift enabled, we reduce the
        azimuth throw required from 40.8$^{\circ}$ to 27$^{\circ}$,
        thereby increasing the density of the planet scan. As a
        result, we enable drift mode for all planet observations.
\end{itemize}

When scheduling multiple calibration targets within one observing
session, we at times run into schedule conflicts. Due to the nature of
the calibration scan, we avoid trimming a planet scan unnecessarily
and favor full coverage of the target wafer group whenever possible.
Therefore the scheduler will choose the target with higher priority
and discard the overlapping lower-priority target when conflict
occurs. To achieve balanced statistics of different calibration
targets while obeying provided priority, we use a round-robin
scheduling policy for calibration targets:
\begin{itemize}
  \item We first compute all scan options for each calibration target.
        Each day we expect to have two options for each target, one
        when the target is rising in the sky and one when it's
        setting.
  \item We then iterate through the list of calibration targets
        following the predetermined order. For each calibration
        target, we iterate through the list of the scan options for
        this target and inject into the master observing plan the
        first scan option that doesn't have a conflicting calibration
        scan. This scan option will also be removed from the option
        list. After we exhaust the list of scan options for a target,
        we discard this target from the target list, and when we have
        iterated through all calibration targets while there are still
        unscheduled scan options, we reiterate from the beginning of
        the list, repeating the same steps until all options are
        either scheduled or dropped due to scheduling conflicts.
  \item When the calibration scan conflicts with the baseline CMB
        scans, on the other hand, we simply trim the CMB scans to make
        space for calibration scans, as at the early stage of SO we
        chose to prioritize better characterization of the instrument
        over maximizing the science outputs.
\end{itemize}
The effect of this round-robin scheduling policy is that calibration
targets that have discarded a scan option due to schedule conflict
will have higher priority in scheduling its next scan option.

\subsection{Sun avoidance}\label{sec:sun avoidance}
Direct exposure to sunlight may severely damage our instrument. The
scheduler ensures the observing plan is Sun-safe with the following
steps.
\begin{itemize}
  \item For each scan block, we use the start time of the block as a
        reference time and compute the sky coordinates of the Sun in
        equatorial coordinates using
        \texttt{pyephem}.\footnote{\url{https://rhodesmill.org/pyephem/}}
  \item We create an empty map in equatorial coordinates covering all
        possible right ascension and declination reachable by SO. For
        each pixel of the map, we compute the distance between this
        pixel and the Sun and store it as a sun distance map.
  \item We identify pixels that fall within a predefined exclusion
        radius (41$^\circ$ for SAT) and mask them as Sun-unsafe.
        Knowing a pixel is Sun-safe at a particular reference time is
        insufficient as the Sun also drifts on the sky. To calculate
        how long a pixel will remain Sun-safe, we use the fact that
        the Sun moves effectively along the axis of right ascension at
        0.25 arcminute per second. For a pixel that is 1$^\circ$ away
        from the nearest Sun-exposed pixel at the same declination,
        this pixel will remain safe for 240 seconds before entering
        the exclusion radius from the Sun. We define this duration as
        the Sun-safe duration.
  \item We compute the Sun-safe duration for all the pixels, similar
        to the Sun distance map. For a given block we then compute all
        pixels that will be covered by the scan, and if any pixel has
        a Sun-safe duration less than $d+30$ minutes, with $d$ the
        duration of the block in minutes, we flag this block as
        Sun-unsafe and trim it such that no pixels will violate the
        Sun-avoidance rule. The 30 minutes interval is chosen
        specifically as a last-resort response time such that when all
        Sun avoidance procedures have failed, the on-site team will
        have this amount of time to manually protect the instrument.
\end{itemize}

\subsection{Operation planning}\label{sec:operation}
With an observing plan established, we need to plan for the time
needed for various instrumentation tasks and fully account for the
state progression in the schedule. This is especially important for
planet scans as observing a planet passage at a target elevation
requires an accurate pre-planning to prepare the telescope. As an
example, during the early commissioning stage, SATs have the
requirement that the HWP should be spinning during a planet scan; we
also require the HWP to stop spinning when the telescope elevation is
changed to protect the HWP. Suppose spinning up and down the HWP both
take 20 minutes. In order to start a planet scan at time $t_0$, the
minimum time required to prepare the telescope in the required state
depends on the state of the system. As illustrated in Table~\ref{tab:
  hwp time}, depending on whether the HWP is already spinning and
whether an elevation change is necessary, the minimally required setup
time before the planet scan varies significantly. HWP operation is one
of the many such operations that has a dynamic setup time based on the
prior state. It is thus crucial for the scheduler to fully track the
state progression during an observing session to observe a planet at
the scheduled time.

\begin{table}[h]
  \centering
\caption{An example of how the detector setup time depends on the state of the telescope.}
\begin{tabular}{|c|c|c|c|}
\hline
Case & HWP Status & Elevation Change & Setup Time \\
\hline
1 & Spinning & No & 0 minutes \\
\hline
2 & Not Spinning & No & 20 minutes \\
\hline
3 & Spinning & Yes & 40 minutes \\
\hline
4 & Not Spinning & Yes & 20 minutes \\
\hline
\end{tabular}
\label{tab: hwp time}
\end{table}

To track state dependency and the associated time cost for different
operations, we define each operation as a function that takes in the
observatory state and returns a modified state together with the
duration of the operation (i.e., \texttt{operation(state) ->
  (new\_state, duration)}). We group operations based on how they are
scheduled.
Pre-session operations are observatory setup operations scheduled at the beginning of an observing session, which typically lasts a day. Post-session operations are wrap-up operations scheduled at the end of the session. Pre-scan operations include setup tasks performed before each scan block, while post-scan operations are scheduled right after each scan block.

Given the state of the instrument at $t_i$ as $s_i$, applying an
operation, \texttt{op}$_i$, will progress the state to $s_{i+1}$ with
$(s_{i+1}, T_i) = \mathtt{op}_i(s_i)$ and increment the clock with
$t_{i+1} = t_i + T_i$, where $T_i$ is the duration of operation $i$. A
basic requirement for scheduling a state-dependent operation is that
\texttt{op}$_i$ cannot be scheduled before $t_i$, otherwise the state
$s_i$ will no longer be valid. This poses an important causal time
constraint during operation scheduling. This constraint requires a
strict chronological state progression, and thus leaves little room
for prioritization. On the other hand, in practice we often want to
prioritize some scans over the others, such as prioritizing
calibration scans over the baseline CMB scans in the early stage of
SO. To solve this problem we use a multi-pass scheduling approach. The
basic idea of this approach is that in the first pass we allow
calibration operations to violate the casual constraint and extend
into previous non-calibration operations. The invalid state will be
corrected in the second pass. In the second pass we first trim the CMB
operations that overlap with calibration operations, and then
re-schedule all operations with causal constraint to produce an
operation plan with valid state progression. A scan that cannot fit
into the constrained time window will be skipped.

The detailed steps in the scheduler are as follows:
\begin{enumerate}
  \item \textbf{Initializing the scheduler}: At time $t_0$, the
        scheduler determines the initial state of the instrument,
        $s_0$. It then adds pre-session operations at $t_0$ and
        progresses the state and clock to $s_1$ and $t_1$,
        respectively.
  \item \textbf{First-pass planning}: Starting at $t_1$ with state
        $s_1$, we perform the first pass planning by iterating through
        the list of scans in the observing plan in chronological
        order. For each scan block, we define a time constraint that
        specifies the time window in which all operations for the scan
        should be scheduled. For baseline CMB scans, we use the causal
        time constraint, and for a calibration scan, we extend the
        time constraint to whenever the previous calibration operation
        is scheduled, to avoid conflicting with other calibration
        operations. We add operations for a scan to the schedule in
        the order of pre-scan operations, scan operations, and
        post-scan operations, progressing the state and clock
        accordingly.
  \item \textbf{Optimizing scan operations}: When adding operations
        for a scan within the time constraint, if the duration of the
        pre-scan operations is short enough that they finish well
        before the scheduled time, we update the time constraint based
        on the estimated gap time, backtrack the state to before the
        operation, and re-inject the operation to the plan to avoid
        unnecessary gaps between pre-scan operations and the actual
        scan operation. In the case that the time constraint does not
        leave enough time for the pre-scan operations to complete, the
        scan block will be trimmed to the current clock time. On the
        other hand, when scheduling a post-scan operation and if the
        time constraint leaves not enough time to complete the
        operation, the scan block will be trimmed by the expected
        excess time, and we will backtrack the state to re-plan the
        block. Figure~\ref{fig:constraint} illustrates the effect of
        this constrained planning, where the scan block has to be
        trimmed to make time for pre- and post-scan operations within
        time constraint.
  \begin{figure}[t]
    \centering
    \includegraphics[width=0.8\textwidth]{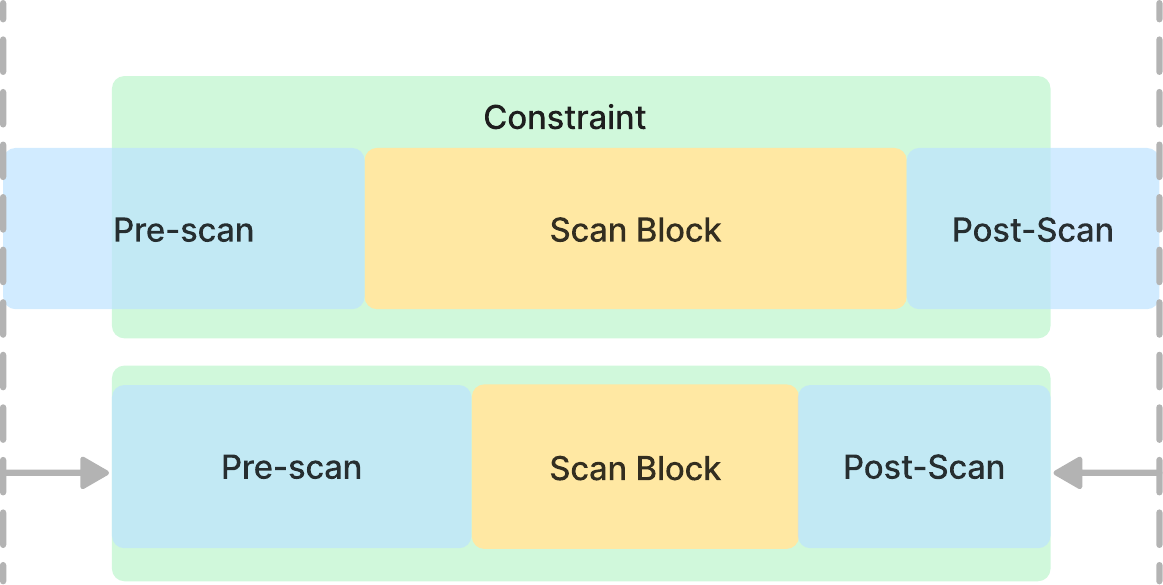}
    \caption{Constrained planning of pre-/post-scan operations.}
    \label{fig:constraint}
  \end{figure}
  \item \textbf{Second-pass planning}: In the second pass, we repeat
        the same steps as the first pass with some important
        differences in how the time constraint for a scan is set.
        Specifically, in the second pass, we set the time constraint
        for CMB scans such that they obey the causal constraint and
        also do not run into any calibration operations in the future,
        based on the result of the first pass. For calibration scans,
        we require simply a causal constraint in the second pass. This
        second pass will produce an operation plan with valid state
        progression.
\end{enumerate}

\subsection{Azimuth movements}\label{sec:azimuth}
During operation planning, we must ensure that both pre-scan and
post-scan operations are safe from direct sunlight exposure. Instead
of restricting the time constraints to a sun-safe time window, which
can lead to loss of observation time, we employ an alternative
approach. While our instrument has strict elevation requirement for
pre-scan setup operations, we are flexible to move in azimuth. We can
therefore find an azimuth ``parking'' spot that is sun-safe for the
duration of the setup and perform the operations there before moving
back to the target azimuth.

For instance, if a particular pointing is sun-safe for 10 minutes, but
the expected duration of the pre-scan operations is 20 minutes, we
will not be able to observe this target at all if we restrict the
pre-scan operations to complete within the sun-safe time.
Alternatively, we can find an azimuth that is sun-safe for over 20
minutes, move to this position, perform the operations, and then move
back to the target azimuth without losing any observation opportunity.
The SO scheduler follows this alternative approach when scheduling
pre-scanning operations. Post-scanning at the time of writing is
typically instantaneous so we do not move the telescope unnecessarily
and restrict the operations to complete within the sun-safe interval
of the original scan.

\begin{figure}[t]
\centering
\includegraphics[width=0.8\textwidth]{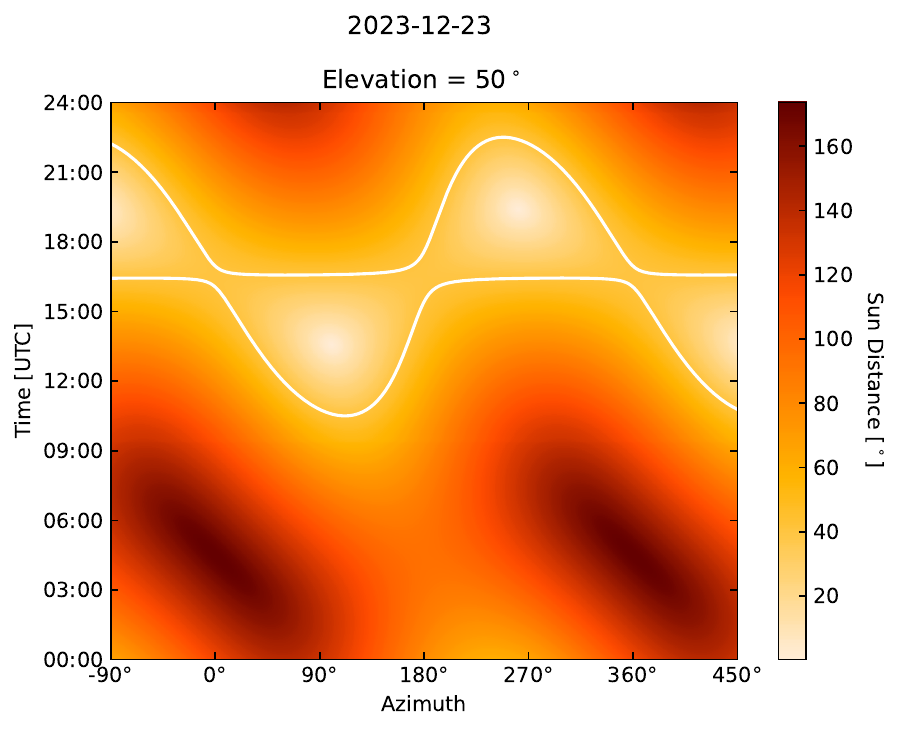}
\caption{Distance to the sun at various times and azimuths on December 23th, 2023, at 50$^\circ$ elevation. The contour indicates the sun exclusion radius of $41^\circ$.}
\label{fig:sun avoidance}
\end{figure}

SO platforms can move azimuthally from $-90^\circ$ to 450$^\circ$.
When moving between two points our platform may choose either a
clock-wise or counter-clockwise move. During daytime observations, one
of the two moves may encounter the Sun in the path.
Figure~\ref{fig:sun avoidance} shows an example of how sun distance
changes as a function of time and azimuth at 50$^\circ$ elevation. One
can easily see that if when moving from 180$^\circ$ to 0$^\circ$ at
around 12:00 UTC, we may reach a sun-exclusion zone moving
counter-clockwise, whereas if we clockwisely move from 180$^\circ$ to
360$^\circ$, the move is permitted by sun avoidance rule. One also
sees that at around 17:00 no azimuth no sun-safe azimuth can be found
when observing at this elevation. In cases like this we may either
choose to observe at a different elevation, or abort the schedule and
stow our telescope.

To find the azimuth move sequence that never reaches a sun-exclusion
zone, we consider all possible azimuth move patterns and find an
optimized sequence that minimally deviates from the original plan,
while ensuring sun-safety of operations and avoiding any motion that
would encounter the Sun.

Azimuth planning follows operation planning and involves optimizing
the sequence of azimuth moves. We start by iterating through the list
of operations. For each pre-scan operation, we calculate the azimuthal
ranges that are sun-safe for the duration of the operation. Instead of
searching through a fine grid of azimuths, we select a few special
values as options. For a pre-scan operation, we consider the
boundaries of the azimuth ranges as potential options. If the
operation is sun-safe for its duration, we also consider all possible
angle unwrappings of the original pointing within the instrument's
limit, ($-90^\circ$ to 450$^\circ$) as options. We assume operations
during the actual scan are sun-safe, ensured by previous sun avoidance
rule, and consider valid unwrappings of the original pointing as valid
options. The reason of these particular azimuth choices will become
clear later.

Let az$_i^{(0)}$ denote the original pointing for operation $i$, and
let az$_i^{(j)}$ denote the $j$-th azimuth option for this operation.
We define the best azimuth choice, $\tilde{\rm az}_i$, for operation
$i$ as the one that minimizes the objective function:
$$\tilde{\rm az}_i = \arg\min_{\mathrm{az}_i^{(j)}} \sum_i \left\vert{\rm mod}\left[(\mathrm{az}_i^{(j)} - \mathrm{az}_{i}^{(0)}) + \pi, 2\pi\right] - \pi \right\vert$$

This problem has the structure of a linear programming problem, so we
expect the optimal solution to exist at one of the vertices of the
feasible region. This is why we consider only a discrete set of
azimuth options. To find the optimal azimuth sequence, we recursively
explore all possible choices of azimuth for each operation, discarding
any path that would encounter the Sun. We then identify the sequence
that minimizes the objective function. If such a sequence is found, we
assign the resulting azimuth to each operation accordingly. If no safe
sequence is found, the schedule reports a failure, indicating to the
remote observer that the schedule should be truncated or skipped to
avoid the Sun.

\section{Automated Workflow System and Automated Data Processing}
\label{sec:workflow}
The massive data volume of SO motivates the development of an
automated and scalable data processing pipeline. SO has adopted a
modern workflow management system, \texttt{prefect}, which is
open-source software that offers a robust infrastructure to manage
data pipelines at scale. Unlike traditional approaches such as
\texttt{cronjobs}, \texttt{prefect} offers both transparency via
online monitoring of running status and logs, and flexibility in
managing complex dependencies as well as concurrency requirements,
making it an ideal framework to coordinate the automated data
pipelines in SO.

The daily data reduction pipeline in SO is modularized into a series
of executable scripts with inter-dependencies. To simplify these
dependencies, we follow a design choice that each pipeline module
should be independently executable and should maintain its own
registry of processed datasets and the corresponding data products. We
refer to this registry as a manifest database (ManifestDB). Most
pipeline scripts that operate over a list of observations maintain
their own ManifestDB, which enables the decoupling of complex
interdependencies between pipeline modules. As a result, each module
can be executed independently in any order, with dependencies resolved
through queries to its prerequisite modules' ManifestDBs.

\texttt{Prefect} introduces the concept of a \texttt{Flow}, which
represents a self-contained, executable unit within a pipeline. This
concept integrates with the modular design of SO's data processing
pipeline. Moreover, \texttt{Prefect} defines a \texttt{Deployment} as
a \texttt{Flow} deployed with specific settings. In SO, we wrap each
data processing pipeline module as a distinct \texttt{Flow} and create
a separate \texttt{Deployment} for each telescope, accommodating
potential variations in instrument configurations as well as storage
locations.

The \texttt{Prefect} workflow system allows for the separation of the
host environment from the execution environment of flows, enabling
remote deployment of flows over a network. In this remote deployment
model, flow runs are submitted to a work pool, which distributes the
runs among subscribed workers. SO operates two work pools, each
responsible for a distinct part of the data processing pipelines. A
dedicated compute node is utilized for automated data processing,
where pipeline workers are deployed and subscribed to the
corresponding work pool. To manage software dependencies, each worker
is containerized and deployed using a Docker
image\cite{software:docker}, which provides the necessary software
dependencies. A key requirement for scalable data pipelines is
concurrency management. To address this, we leverage another
\texttt{Prefect} feature called \texttt{Queue}. Flows that modify the
same database are placed on a queue with a concurrency limit of 1,
ensuring that data inconsistencies arising from concurrent writes are
avoided.

This workflow system is deployed in SO to orchestrate the automated
data processing pipelines at the site, which encompasses the critical
steps that transform ephemeral data from the detector readout systems,
SLAC Microresonator RF (SMuRF) Electronics\cite{Henderson:2018,Yu:2023} (referred as
SMuRF servers hereafter), into permanent archival data format known as
\texttt{Books}, perform necessary preprocessing, and, eventually,
produce maps of the CMB sky on a routine basis. This full pipeline can
be divided into two main stages: data packaging and data reduction.

\subsection{Data packaging}
The first main stage of the automated data processing pipeline is data
packaging. This step is responsible for aggregating and re-bundling
raw timestreams from different SMuRF servers and shipping them to
long-term data storage sites. A full description of the data packaging
is beyond the scope of this paper. Here we provide only a brief
overview.

\begin{figure}[t]
  \centering
  \includegraphics[width=\textwidth]{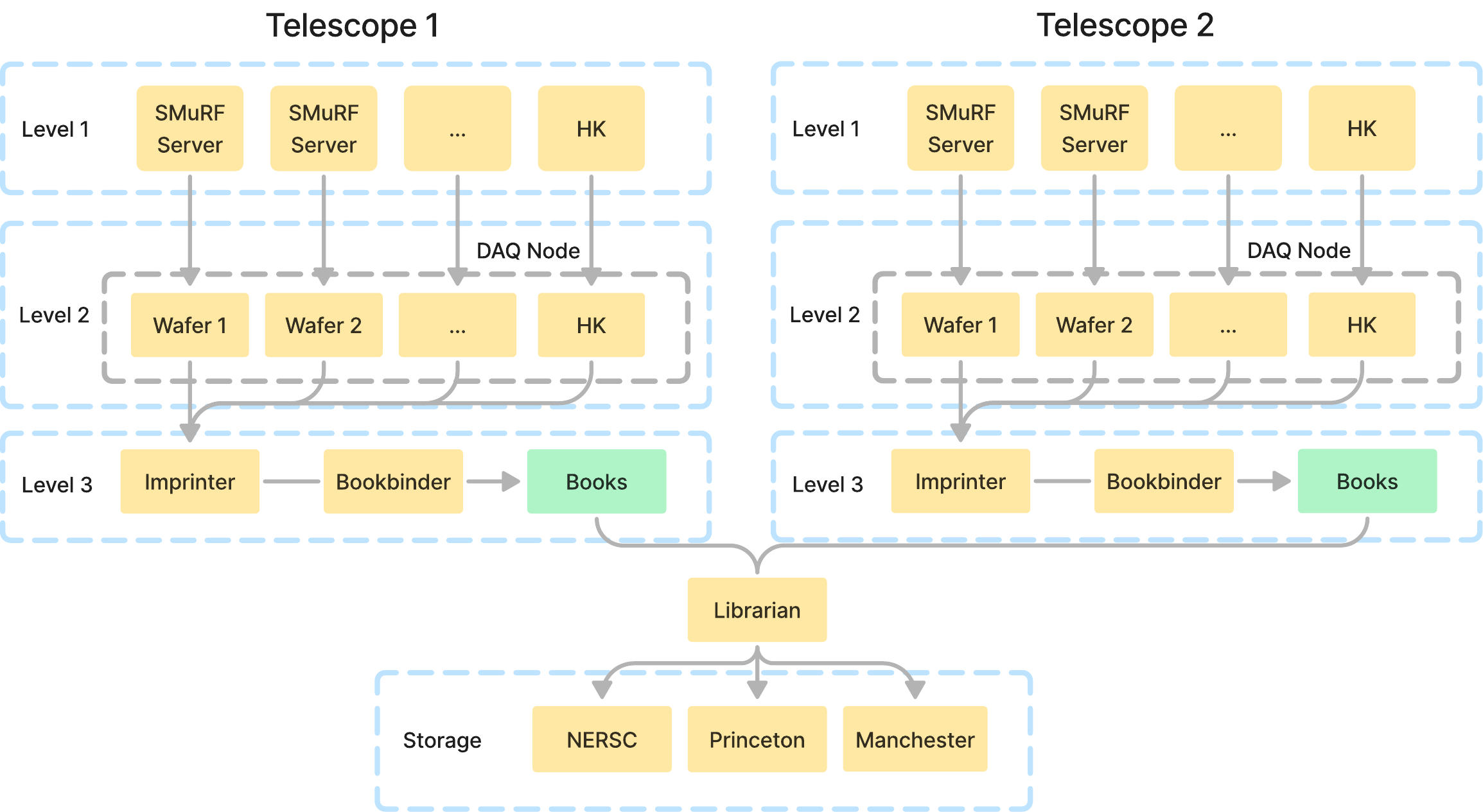}
  \caption{Overview of data packaging pipeline}
  \label{fig:dp overview}
\end{figure}

SO employs a three-level process to describe the different life stages
of our data, as shown in Figure~\ref{fig:dp overview}. This process
begins with the collection of data by the readout systems. This
initial data are saved directly to the SMuRF servers to prevent
network outages from causing permanent loss of data. We refer to the
data at this stage as level-1 data. Each SO telescope has multiple
SMuRF servers so the level-1 detector data are saved across many
machines. In the second stage, the level-1 data are aggregated onto
the data acquisition (DAQ) nodes for temporary storage. We refer to
these aggregated data as level-2. Data at this stage can be broadly
classified into two types: SMuRF data and housekeeping (HK) data, with
the former accounting for the raw detector timestreams and the latter
including ancillary data from supporting hardware such as telescope
pointing information, receiver temperatures and pressures, and so on.

Level-2 data are organized and tracked on a per-wafer basis based on the actions commanded to SMuRF (e.g., ``stream data'', ``take IV curve''), and are stored across multiple files in G3 format, where G3 is a serialized data format developed by SPT-3G\cite{Sobrin:2022} adopted as the main data container format for SO.\footnote{\url{https://github.com/CMB-S4/spt3g_software}} We register these ``level-2 observations'' and their associated G3 files in an SQLite database called \texttt{G3tSmurf} DB and similarly register the housekeeping (HK) data in a database called \texttt{G3tHK} DB. These registries allow for efficient compilation of the cross-wafer observations that correspond to the same physical scan, along with the relevant housekeeping data produced during the same session. This compilation of cross-wafer observations is implemented by a tool called \texttt{Imprinter}.

In particular, \texttt{Imprinter} registers each set of cross-wafer
observations in its own database and initiates the \texttt{Bookbinder}
process that performs the actual aggregation and re-bundling of G3
files from the multi-wafer observations into a compact format known as
a \texttt{Book}. Each G3 file is composed of G3 frames which represent
the smallest data storage units during reading or writing. Each G3
frame can contain an arbitrarily defined subset of the detector
timestreams during data acquisition. The bookbinder also re-divides
the G3 frames based on scan patterns such that each frame contains
data from a single azimuth sweep. This makes each frame more
physically interpretable and more natural to analyze. In addition,
\texttt{Bookbinder} also catches common data processing failure modes,
such as missing samples, missing pointing information, or timing
system failures. We refer to the \texttt{Books} as level-3 data and
consider them to be permanent and immutable data products for
long-term storage and data analysis. SO defines five types of
\texttt{Books} as listed in Table~\ref{tab:book types}.

\begin{table}[h]
\centering
\caption{Types of \texttt{Books} in the Data Packaging System}
\begin{tabular}{|l|p{10cm}|}
\hline
\textbf{Book Type} & \textbf{Description} \\
\hline
\texttt{Observation} & Books containing actual measurements of the sky. \\
\hline
\texttt{Operation} & Books recorded during detector operations such as IV measurements or bias steps, containing relevant information for detector characterization. \\
\hline
\texttt{Housekeeping} & Books containing data recorded from the housekeeping feeds. \\
\hline
\texttt{Stray} & A catch-all category for level 2 detector timestreams that do not fit into the above categories. \\\hline
\texttt{Smurf} & The calibration and metadata output from the various Smurf systems during observations and operations. \\
\hline
\end{tabular}
\label{tab:book types}
\end{table}
Once bound, books are registered and uploaded to the
\texttt{Librarian} system \cite{Librarian}, which serves as both a
bookkeeper and a transport layer, synchronizing the books across all
of our data storage sites, including the National Energy Research
Scientific Computing Center (NERSC), the Princeton Research Computing
Center, and the University of
Manchester. 

This data packaging pipeline is coordinated by the \texttt{Prefect}
workflow system to run automatically. In particular, the pipeline is
broken down into five flows, as listed in Table~\ref{tab:data
  packaging flows}, each of which is an independently executable
pipeline module responsible for one step in the data packaging
pipeline. For example, \texttt{update-g3tsmurf-db} and
\texttt{update-g3tsmurf-hk} flows are responsible for registration of
level-2 SMuRF data and housekeeping data, respectively.
\texttt{update-book-plan} flow implements the \texttt{imprinter} and
defines the plans for bookbinding, followed by the \texttt{make-book}
flow which triggers the bookbinder to produce the level-3 books. The
last flow, \texttt{upload-book}, calls \texttt{Librarian} service for
book curation and synchronization to remote storage sites.
\begin{table}[h]\centering
\caption{Key data packaging flows managed by the \texttt{prefect} automation system.}
\begin{tabular}{|l|l|}
\hline
Module Name & Description \\
\hline
\texttt{update-g3tsmurf-db} & Registration of level-2 SMuRF Observations and the associated \\ & G3 files in a local \texttt{g3tsmurf} database.\\
\hline
\texttt{update-g3thk-db} & Registration of housekeeping data in a local \texttt{g3thk} database. \\
\hline
\texttt{update-book-plan} & Definition of book as multi-wafer observation set and register\\ & under \texttt{imprinter} database.\\
\hline
\texttt{make-book} & Perform bookbinding based on the definition of books in the \\&\texttt{imprinter} database.\\
\hline
\texttt{upload-to-librarian} & Registration and uploading of books to the \texttt{Librarian} system, \\&which synchronize across storage sites.\\
\hline
\end{tabular}
\label{tab:data packaging flows}
\end{table}

\subsection{Daily Data Reduction}
The second main stage of the automated pipeline in SO is the automated
data reduction pipeline, which begins where data packaging ends. Using
level-3 books as primary inputs, this pipeline executes a series of
data reduction steps, such as detector calibration, matching each
detector with its observed resonator frequency (a requirement of the
microwave-multiplexing system in SO\cite{McCarrick:2021}), solving for
the HWP angle solution using HWP encoder data, and preprocessing each
observation to flag glitches, jumps, and cut bad detectors. These
steps are necessary for achieving daily mapmaking, a crucial target of
the data reduction pipeline and a key milestone for time-domain
astronomy. Table~\ref{tab:site pipeline scripts} lists some example
data reduction pipeline modules deployed during commissioning.
Additional modules, including daily mapmaking, will be integrated in
this pipeline in the future.

\begin{table}[h]\centering 
\caption{List of data reduction pipeline modules deployed during commissioning. Additional modules including daily mapmaking will be integrated in the future.}
\begin{tabular}{|l|p{10cm}|}
\hline
Step & Description \\
\hline
\texttt{update-obs-db} & Registration of observations in a database for subsequent pipeline. \\
\hline
\texttt{make-hwp-solutions} & Solve HWP angle solutions. \\
\hline
\texttt{update-det-match} & Match detectors with the observed resonator frequency. \\
\hline
\texttt{update-smurf-caldbs} & Perform detector calibration and store results into calibration databases (CalDBs). \\
\hline
\texttt{preprocess-tod} & Perform preprocessing steps on each \textit{Book}, including flagging jumps, glitches, bad detectors, and more. \\
\hline
\end{tabular}
\label{tab:site pipeline scripts}
\end{table}

\section{Conclusion}
\label{sec:conclusion}
SO requires optimized observation scheduling to maximize its
scientific impacts. This paper presented an overview of the
multi-level scheduling process in SO. Optimized scan strategies are
designed for each telescope at the highest level to maximize their
respective science goals. These scan strategies are then transformed
into practical observing scripts that operate the telescope, taking
into account detailed instrumentation requirements, such as the
injection of calibration scans and additional observing constraints,
like sun avoidance rules. The observing plan is subsequently
transformed into an operational plan, which accounts for state
progression during operations and the expected time cost of each
operation. This optimized multi-level scheduling process enables
effective realization of the optimized scan strategy while adapting to
instrumentation requirements during day-to-day operations.

SO will collect a vast amount of data to achieve the highest precision
CMB maps. We need both scalable and automated data processing pipeline
at the site to effectively handle this large volume of data. The data
processing pipeline at the site involves both data packaging and daily
data reduction, starting from the aggregation of ephemeral data from
the readout servers up to the daily production of maps. In this paper,
we described an automated workflow system deployed at the site, which
coordinates the automated data pipelines at the site. This workflow
system, based on the open-source \texttt{prefect} system, offers
several advantages over traditional tools like \texttt{cronjobs}, such
as transparency of running jobs, flexibility of deployment, and
concurrency handling. We have successfully applied this system to
coordinate automated data packaging at the site, including the
re-bundling of multi-wafer observations into compact \texttt{Books}
and transporting them to external storage sites. Additionally, this
workflow system also manages the daily data reduction pipeline, which
includes daily generation of data cuts, jump fixes, and HWP solutions,
with the goal of automating daily mapmaking in the near future.

\acknowledgments
This work was supported in part by a grant from the
Simons Foundation (Award \#457687, B.K.). YG acknowledges the support
from the Dunlap Institute for the duration of this work. The Dunlap
Institute is funded through an endowment established by the David
Dunlap family and the University of Toronto. ADH acknowledges support
from the Sutton Family Chair in Science, Christianity and Cultures,
from the Faculty of Arts and Science, University of Toronto, and from
the Natural Sciences and Engineering Research Council of Canada
(NSERC) [RGPIN-2023-05014, DGECR-2023-00180]. We acknowledge various
open source software packages, including \texttt{numpy},
\texttt{pyephem}, \texttt{jax}, \texttt{prefect}, which have been
instrumental in this work.

\bibliography{report}
\bibliographystyle{spiebib}

\end{document}